# EXPLORING INTERMEDIATE (5–40 AU) SCALES AROUND AB AURIGAE WITH THE PALOMAR FIBER NULLER

J. Kühn[1], B. Mennesson[1], K. Liewer[1], S. Martin[1], F. Loya[1], R. Millan-gabet[2], and E. Serabyn[1]

[1] Jet Propulsion Laboratory, California Institute of Technology, 4800 Oak Grove Drive, Pasadena, CA 91109, USA; jonas.kuehn@a3.epfl.ch
[2] NASA Exoplanet Science Institute, California Institute of Technology, Pasadena, CA 91125, USA

## ABSTRACT

We report on recent $K_s$-band interferometric observations of the young pre-main-sequence star AB Aurigae obtained with the Palomar Fiber Nuller (PFN). Reaching a contrast of a few $10^{-4}$ inside a field of view extending from 35 to 275 mas (5–40 AU at AB Aur's distance), the PFN is able to explore angular scales that are intermediate between those accessed by coronagraphic imaging and long baseline interferometry. This intermediate region is of special interest given that many young stellar objects are believed to harbor extended halos at such angular scales. Using destructive interference (nulling) between two sub-apertures of the Palomar 200 inch telescope and rotating the telescope pupil, we measured a resolved circumstellar excess at all probed azimuth angles. The astrophysical null measured over the full rotation is fairly constant, with a mean value of 1.52%, and a slight additional azimuthal modulation of ±0.2%. The isotropic astrophysical null is indicative of circumstellar emission dominated by an azimuthally extended source, possibly a halo, or one or more rings of dust, accounting for several percent of the total $K_s$-band flux. The modest azimuthal variation may be explained by some skewness or anisotropy of the spatially extended source, e.g., an elliptical or spiral geometry, or clumping, but it could also be due to the presence of a point source located at a separation of ∼120 mas (17 AU) with ∼$6 \times 10^{-3}$ of the stellar flux. We combine our results with previous Infrared Optical Telescope Array observations of AB Aur at H band, and demonstrate that a dust ring located at ∼30 mas (4.3 AU) represents the best-fitting model to explain both sets of visibilities. We are also able to test a few previously hypothesized models of the incoherent component evident at longer interferometric baselines.

## 1. INTRODUCTION

Planet formation and migration processes can currently be studied through the direct observation of planets with masses of tens of $M_J$ at large distances (tens of AU) from young (<10 Myr) stars. At 144 pc from us, the young (4 ± 1 Myr) massive ($M \sim 2.4 \pm 0.2\,M_\odot$; DeWarf et al. 2003) pre-main-sequence star AB Aurigae (AB Aur) presents as an ideal prototype. As a member of the Herbig Ae class, AB Aur has been extensively investigated in the near-infrared (NIR), notably by direct and polarimetric imaging of the outer cold disk, which extends out to hundreds of AU. In general, the AB Aur disk is thought to be seen nearly face-on, seen with an inclination at or below 30° (Eisner et al. 2003; Fukagawa et al. 2004). The cold outer gaseous nebula (Grady et al. 1999) has been reported to exhibit spiral geometry with several trailing spiral arms up to 450 AU away from the host star (Fukagawa et al. 2004). In the case of the outer cold dust envelope, it was detected out to 130–140 AU, reportedly presenting both an azimuthal gap at a separation of ∼100 AU, as well as two rings at radial distances of ∼40 and 100 AU, separated by an annular dip (Hashimoto et al. 2011; Oppenheimer et al. 2008; Perrin et al. 2009). There is also evidence of spiral-shaped asymmetry and warped geometry for the closer inner disk structure at a separation of ∼40 AU (Hashimoto et al. 2011).

On the other hand, interferometric investigations (10–350 m baselines) that probe the inner 1 AU, have reported a similarly complex picture very close to the star. Most of the NIR excess was mapped interferometrically and modeled as a ring of hot dust (1800–1900 K) with a sharp inner rim located at 0.2–0.3 AU (Millan-Gabet et al. 1999, 2006b), with a smooth homogeneous hot (2500–3500 K) gaseous contribution inside of it (Tannirkulam et al. 2008). These two inner components are found about equally bright, and together contribute 70% of the total flux at H band (1.65 μm), and 85% at K band (2.2 μm). In addition, closure phase (CP) H-band observations at the Infrared Optical Telescope Array (IOTA) have suggested an off-axis partially resolved feature farther out, at 1–5 AU scales (Millan-Gabet et al. 2006b). Finally, spectral visibilities obtained with the VEGA spectrometer at CHARA in R band have also suggested the presence of an unknown component at about 5.5 AU (38 mas) contributing about 7% of the flux (Rousselet-Perraut et al. 2010). In summary, there is evidence for emission outside a field of view (FOV) of 1 AU, at best only partially resolved with previous interferometric observations. Indeed, this intermediate radial regime, i.e., between the interferometric and single aperture observational regimes, has hitherto not been accessed observationally in the NIR.

Other non-CP interferometric results generally assumed some extended uniform emission (i.e., a "halo") on top of the reported best-fit models, accounting for 5%–10% of the NIR flux, and typically extending from 1.5 to 70 AU (10–500 mas). This extended "halo" component was required to explain the visibility deficit systematically measured at short interferometric baselines (∼10 m; Monnier et al. 2006; Tannirkulam et al. 2008). The spatial extension of such a putative halo was constrained by speckle interferometry observations at K band (Leinert et al. 2001) with visibilities near unity for spatial scales beyond 29 AU (200 mas). In the mid-IR (10 μm), faint halo emission of about 10% of the flux has, however, been reported at larger separation beyond 0″.5 (Mariñas et al. 2006; Monnier et al. 2009), albeit the bulk of the 10 μm emission was measured to be contained within 35 mas (5 AU). Such "halo" features have been reported around young stellar objects (YSOs; Leinert et al. 2001) and FU Orionis objects (Millan-Gabet et al. 2006a), albeit probably with different origins, but are still not well understood.

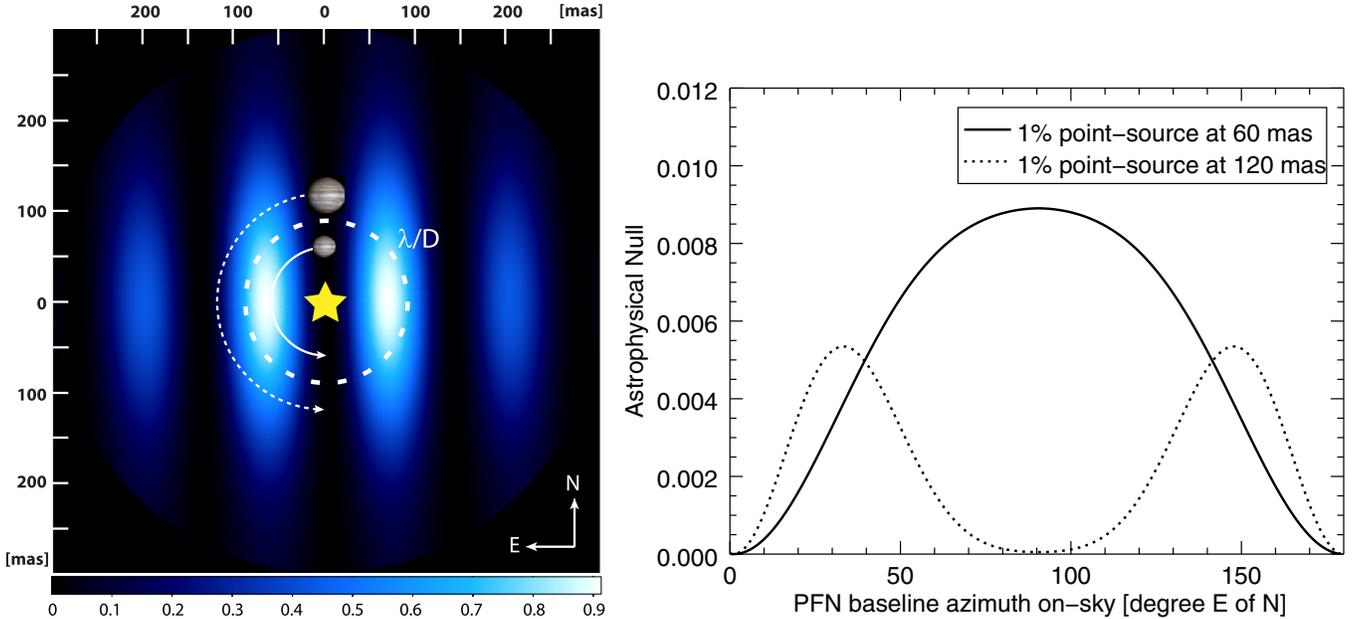

**Figure 1.** Left: PFN transmission function (fringe pattern) on the sky at Palomar. The dashed circle represents a radius of λ/D for the 5.1 m Hale Telescope at $K_s$ band, and the arrows the hypothetic trajectories of potential point-source companions relative to the fringes. The plotted image size is 600 × 600 mas (the map is computed up to 300 mas separation radius). Right: expected astrophysical null modulation that would be observed for these two point sources carrying 1% of the flux of the unresolved star. The azimuthal modulation frequency provides information on separation, while the position angle can be retrieved from the phase offset (modulo 180°).

The AB Aur system presents a complex picture, probably including unseen ongoing planetary formation processes, at a variety of scales. However, the gap of observational coverage in between NIR interferometry and direct imaging lies at critical planet formation scales, ranging from ∼1 to 40 AU. With the Palomar Fiber Nuller (PFN; Martin et al. 2008; Mennesson et al. 2011a, 2010; Serabyn et al. 2010), we can now access most of this scale, i.e., the outer solar system scale past ∼4 AU, in the NIR for the first time. Here we report on $K_s$-band nulling observations of AB Aur, obtained with the PFN 3.2 m interferometric baseline, that cover a 180° range in azimuth. With an effective search space extending from ∼35 mas (5 AU, inner half transmission point) to ∼275 mas (40 AU, outer half transmission point), the PFN is ideally suited to provide measurements at these missing scales.

## 2. OBSERVATIONS AND RESULTS

The PFN is a rotatable-baseline nulling interferometer that makes use of a single-mode (SM) fiber to combine two destructively interfered telescope sub-apertures. All individual beam wavefront phase errors are discarded as transmission losses at the SM fiber tip, and the Palomar P3K adaptive optics (AO) system (Dekany et al. 2013)—acting as a fringe tracker here—maintains the optical path difference (OPD) between the two beams around the central "null fringe" (in good seeing, the OPD rms is typically λ/10). The measured interferometric signal will reach its lowest level (OPD = π) from time to time during a recording sequence (Martin et al. 2008; Mennesson et al. 2011a, 2010). Performance-wise, the PFN delivers very deep on-sky nulls (∼2-4 × $10^{-4}$) on a bright point source at K band, and was previously used to derive some of the best constraints on the presence of a companion within 2 AU of Vega (Mennesson et al. 2011b). The PFN causes the Hale Telescope pupil to be re-imaged and then rotated by means of an actuated K-mirror, which has the effect of rotating an equivalent fixed baseline of $b = 3.2$ m between a pair of elliptical apertures of 3 × 1.5 m. This enables us to explore the close environment of a star and look for azimuthal variations. In the case of an unresolved naked star, or of a uniform disk seen face-on, the measured astrophysical null will show no modulation with baseline rotation, at least down to the current measurement accuracy of a few $10^{-4}$. On the other hand, non-axisymmetric source structure, including the presence and position—*modulo* 180° for the position angle (P.A.) of an off-axis source—will be revealed by a modulation of the null signal as the source flux is modulated by the PFN on-sky transmission (Figure 1).

AB Aur was observed with the PFN in $K_s$ band on the night of 2012 October 30 UT, at six different azimuth baselines spanning 180° on-sky in steps of 30°, all obtained under a seeing of ∼0″.85. Each sequence consisted of 300 s long recordings of 10 ms individual exposures. The observed signal was constantly chopped at ∼12 Hz with a rotating aperture wheel which alternately lets through the following signals (see Figure 2—left): flux from sub-aperture "A", sub-aperture "B", both sub-apertures "A+B", and dark "D" (both beams blocked; Martin et al. 2008). The probability distributions of these different flux signals are the statistically relevant metrics for the astrophysical null, when processed with our null self-calibration algorithm (Hanot et al. 2011; Mennesson et al. 2011a, 2011b), which retrieves the underlying astrophysical null for each recorded on-sky azimuth. This is achieved by comparing the measured A+B null distribution with different synthetic null distributions built from boot-strapped (i.e., resampled) flux samples of A, B, and dark measurements. Each synthetic null distribution depends solely on three parameters (fitted by a Pearson $\chi^2$ minimization): the mean OPD difference between the beams (i.e., the deviation from π), the OPD jitter (rms) around that mean, and the astrophysical null leakage $N_{as}$, which is of course the quantity of interest (Figure 2—right). Such best-fit estimates are obtained for 100 different boot-strapped data sequences, in order to evaluate the error bars on the retrieved astrophysical null. Such astrophysical nulls and error bars are derived for each PFN baseline orientation used.

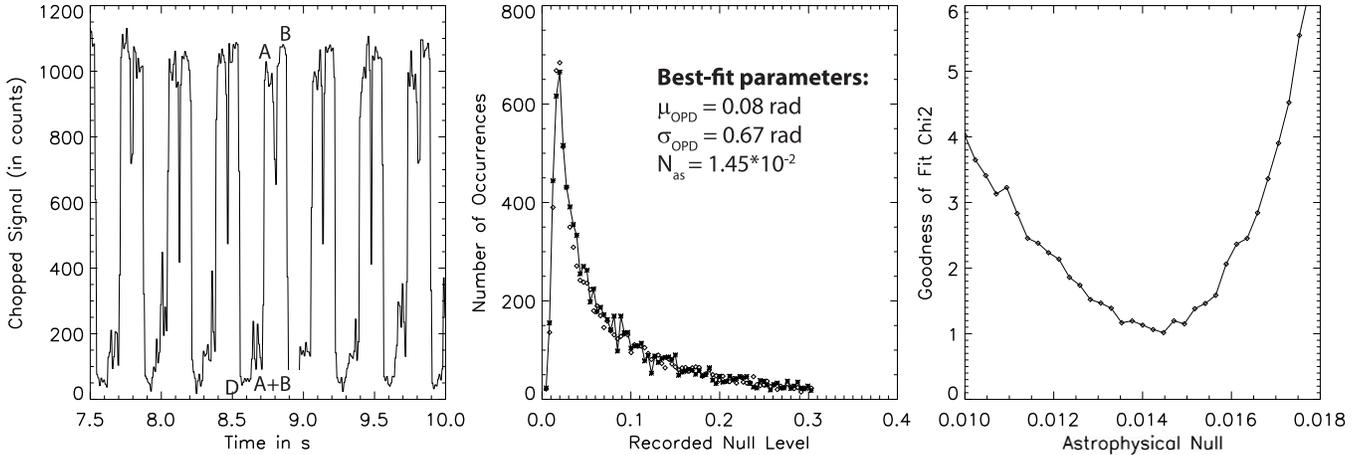

**Figure 2.** PFN on-sky measurements and data reduction. Left: example of recorded raw chopped signal on AB Aur for P.A. = 40° E of N, where a different signal is let through every 80 ms, alternating between dark (D), interferometric null (A + B), and individual signals (A and B) states. Center: example of empirical histogram of recorded nulls (40° P.A., solid line), and one (among 100) bootstrapped best-fit solution (diamonds). Right: goodness of fit estimates (Pearson $\chi^2$ minimization) for a variety of synthetic null distributions built from the recorded A and B signals, as a function of underlying astrophysical null. The best-fit synthetic histogram is obtained for a $1.45 \times 10^{-2}$ null, which is also plotted in the central panel.

**Table 1**
Summary of PFN Observations of AB Aur

| P.A. (deg E of N) | Astrophysical Null $\times 10^{-2}$ | $\sigma_m \times 10^{-2}$ |
|---|---|---|
| 10 | 1.71 | 0.08 |
| 40 | 1.43 | 0.05 |
| 70 | 1.38 | 0.04 |
| 100 | 1.73 | 0.04 |
| 130 | 1.35 | 0.04 |
| 160 | 1.52 | 0.06 |

The resulting astrophysical null leakages measured (abbreviated as "nulls" in the next sections) for AB Aur are plotted in Figure 3 as a function of azimuth, along with their $1\sigma$ error bars $\sigma_m$. For comparison, the figure also displays the instrumental nulls measured on a combination of calibrator stars and the AO internal light source. Subtracting this static instrumental error—nearly an order of magnitude smaller than the measured nulls around AB Aur—provides the calibrated astrophysical nulls[3] plotted in Figure 3 (filled diamond points) and listed in Table 1. It is important to note that any observed astrophysical null represents the amount of circumstellar flux leaking through the PFN fringe pattern (Figure 1) divided by the "central flux" unresolved by the PFN. In the case of AB Aur, this unresolved central component includes both the star *and* any hot dust/gas emission at sub-AU scales (a few mas), as previously reported by Millan-Gabet et al. (1999) and Tannirkulam et al. (2008). This is an important point, as this unresolved hot dust/gas emission is believed to dominate over the K-band starlight by at least a factor of two.

---
[3] While the instrumental null function is flat at the few $10^{-4}$ level under laboratory conditions, a larger residual modulation ($\sim 10^{-3}$) is observed on the sky as the baseline rotates. This is induced by the AO dichroic beamsplitter seen in transmission by the converging science beam, yielding a variable amount of dispersion as the baseline rotates. This effect has been calibrated out, and since these observations, it has been corrected by a compensator plate inserted in the PFN beam.

Two results are evident in Figure 3. First, a fairly constant 1.52% ± 0.05% average null is present at all azimuths. Therefore, the resolved emission detected around AB Aur is dominated by an azimuthally extended source. Second, a slight azimuthal variation on the order of 0.2% ± 0.07% in amplitude is also present, significantly larger than the uncertainties on the retrieved null values. This suggests a small contribution inside the PFN FOV from either some anisotropy in a spatially extended source (e.g., spiral arms or a slight ellipticity), or from a localized off-axis source.

## 3. DISCUSSION AND DERIVED CONSTRAINTS

### 3.1. Extended Emission

Assuming a nearly face-on disk (Eisner et al. 2003; Fukagawa et al. 2004), two simple scenarios can be considered for the dominant spatially extended contribution.

1. A uniform spatially extended "halo" filling the PFN FOV (Figure 4(1)), as suggested by previous speckle (Leinert et al. 2001) and long baseline interferometric observations (Millan-Gabet et al. 2006b; Monnier et al. 2006; Tannirkulam et al. 2008). This halo case is relatively straightforward to analyze, as dividing the measured 1.52% null by the mean PFN transmission over the inner 300 mas (43.2 AU, see Figure 1) yields a total halo flux contributing 6.4% ± 0.02% of the unresolved inner flux, or 6.0% ± 0.02% of the total $K_s$-band flux (Figure 5, dashed–dotted curve). Such an isotropic structure would reproduce the observed average on-sky null (Figure 6, black bold dashed curve) with a $\chi^2$ of 3.7 (Table 2). This rather large ~6% flux could correspond to the "missing" incoherent flux reported with NIR long-baseline interferometry, meaning that most—if not all—of this flux would fall inside the PFN FOV. Indeed, a halo extending further out than the PFN outer-working angle of 40 AU (275 mas) would have to be even brighter to explain the detected leakage. This is in excellent agreement with K-band speckle interferometry observations (Leinert et al. 2001), which exclude significant emission on >29 AU (200 mas) scales (as opposed to mid-IR emission, which is reported to extend beyond 70 AU, Monnier et al. 2009).

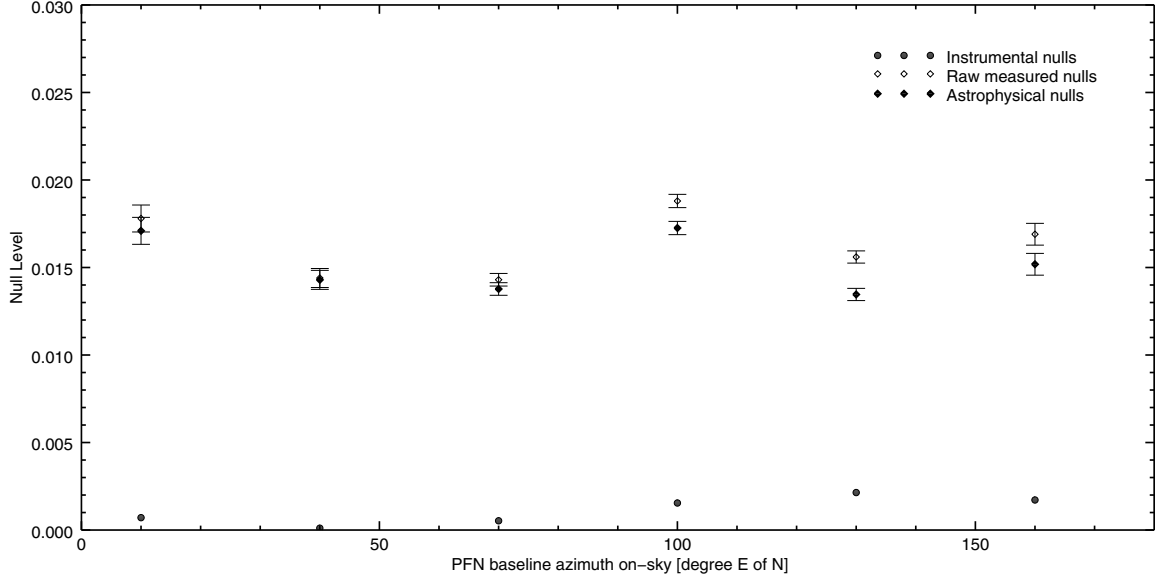

**Figure 3.** AB Aur on-sky nulls recorded with the PFN on 2012 October 30 UT for six different baseline orientations spanning 150° (180° effectively with the π periodicity of the PFN transmission) azimuthal range by increments of 30°. Unfilled diamond points are the raw reduced on-sky nulls. Subtracting the instrumental nulls (filled circles) yields the final calibrated astrophysical null (filled diamonds) presented in Table 1. An average 1.52% null level is observed, with a small ±0.2% azimuthal modulation.

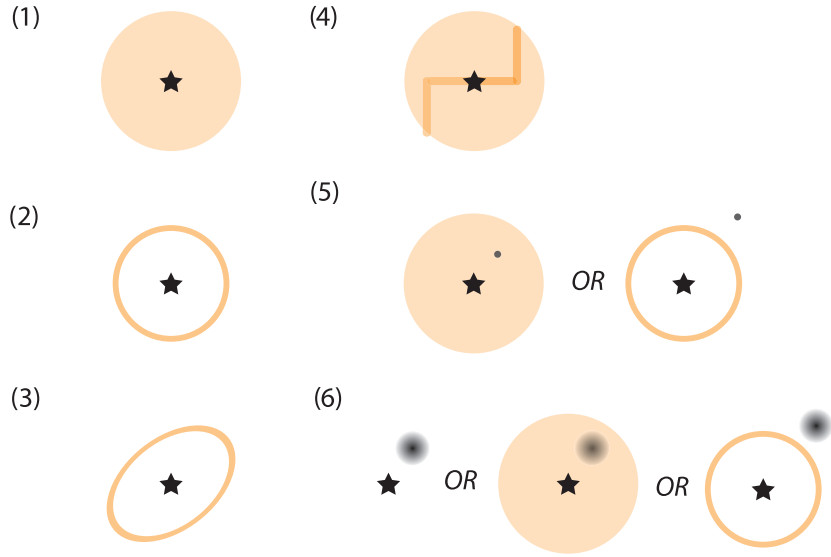

**Figure 4.** Cartoons of the various models considered inside a 30–300 mas FOV to explain the measured PFN astrophysical null response of Figure 3. The star symbol represents the stellar component, including any close-by (<1 AU = 7 mas) NIR emission that is unresolved by the PFN: (1) is a homogeneous extended halo, (2) is a thin ring, (3) is a thin elliptical ring, or a warped circular ring, (4) is a simple model of spiral-shaped emission embedded in an extended halo, (5) is an off-axis point source embedded in an extended halo, or in addition to a thin ring (inside or outside) (6) is an extended Gaussian blob, either alone or embedded in an extended halo, or ring, structure.

**Table 2**
Astrophysical Models Hypothesized or Tested to Explain the Measured PFN Azimuthal Null Response

| No. | Description | Best-fitting or Retained K-flux and Geometric Parameters | $\chi^2$ |
|---|---|---|---|
| 1 | Extended halo | 6% flux within 300 mas | 3.7 |
| 2 | Circular thin ring | 8% flux at 30 mas | 3.7 |
| 3 | Elliptical ring | 6.5% flux, semi major axis ∼200 mas at P.A. = 130°, semi minor axis ∼120 mas | 2.2 |
| 4 | Embedded spiral structure | 1.25% flux embedded in a 4% halo within 300 mas | 2.7 |
| 5 | Embedded companion | 0.62% companion at 200 mas, P.A. = 130° embedded in a 7.1% ring at 30 mas or 5.3% halo within 300 mas | 0.6 |
| 6 | Gaussian hot spot* | 8% blob at 29 mas, P.A. = 12° (2003) with FWHM dia. = 12 mas | 471 |
| 7 | Halo extending further out* | 8% halo within 400 mas | 60 |
| 8 | Unresolved companion* | 2% companion at 9 mas, P.A. = 22° | 395 |
| 9 | 10 μm scaled-down halo* | 5.45% Gaussian halo with FWHM major dia. = 72 mas, P.A. = 74° (2012) and minor dia. = 68 mas | 9.4 |

**Note.** (*) Denotes previous models suggested to fit IOTA (Millan-Gabet et al. 2006b) and Keck segments-tiling (Monnier et al. 2009) observations.

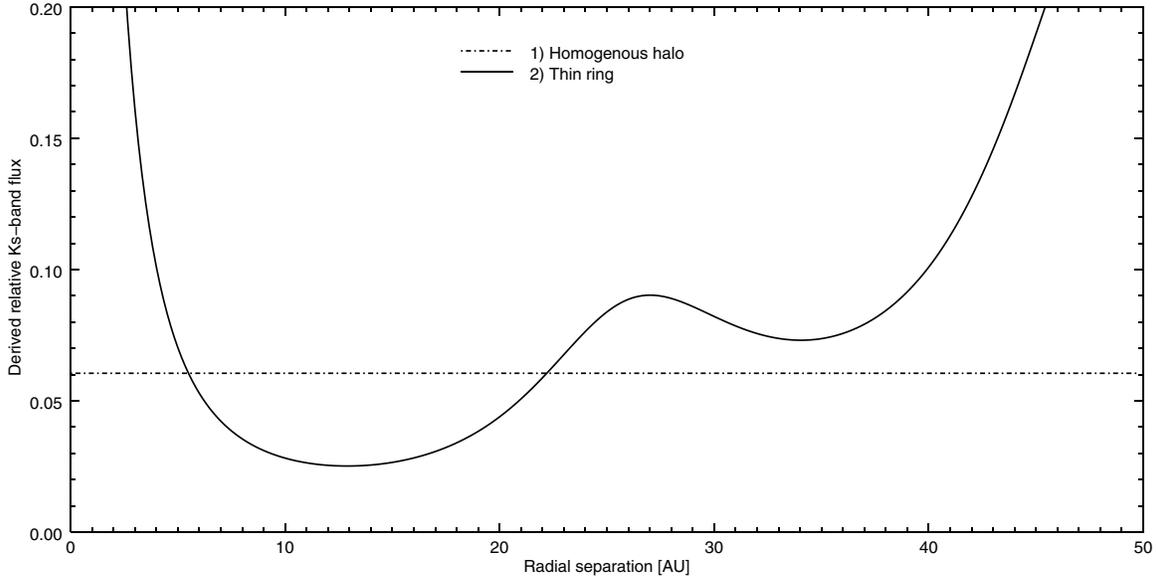

**Figure 5.** Derived constraints on the extended emission responsible for the detected mean null of 1.52%. Realistic flux levels below 10% of the total $K_s$-band flux would place a putative ring (scenario #2) between 3 and 40 AU. An equally valid scenario #1 is a homogenous halo extending over the PFN FOV and contributing about 6% of the total $K$-band flux (dashed dotted line). 1 AU = 7 mas at AB Aur distance.

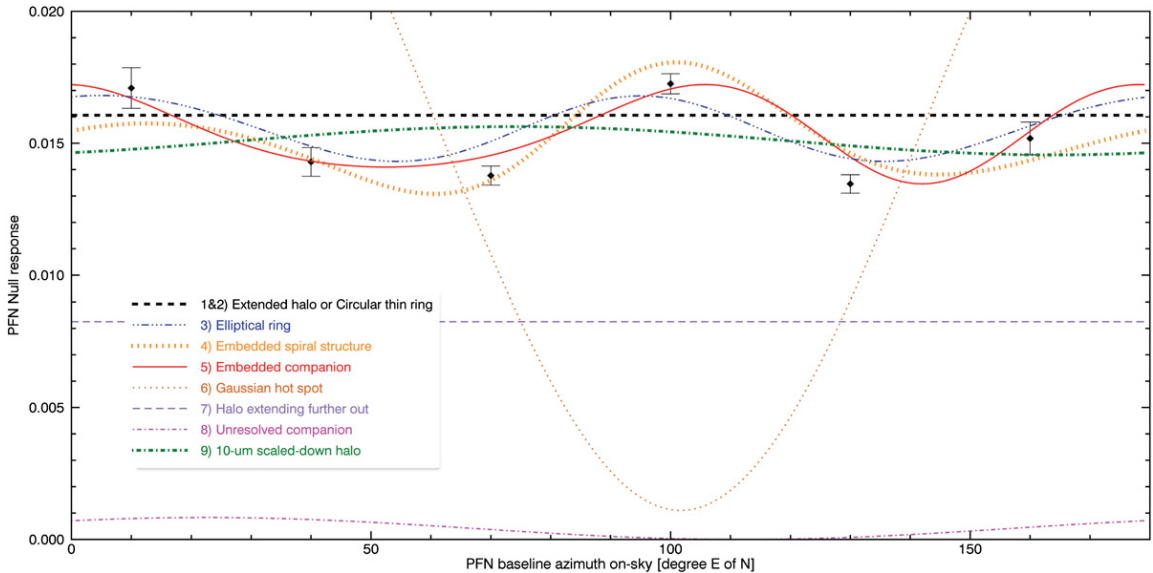

**Figure 6.** PFN measured nulls (diamonds), and best-fitting and previously hypothesized #1–9 models, as listed in Table 2. With just one baseline length (3.2 m) available with PFN, the circularly symmetric cases of a ring, thick annulus, and halo are degenerate.

2. A ring or narrow annulus geometry (Figure 4(2)), though multiple rings are also a possibility. Assuming an arbitrarily thin ring, we can derive the ring flux as a function of its distance from the star (Figure 5, plain curve), that would result in the observed mean null (Figure 6, same black dashed curve for a ring with 8% flux at 4.3 AU) with a $\chi^2$ of 3.7 (Table 2). Given that the unresolved inner component is reported to account for 90%–95% of the total $K$-band flux (Monnier et al. 2006; Tannirkulam et al. 2008), realistic flux levels are 5%–10% max for the missing contribution. Based on Figure 5, this would imply ring separations between 3 and 40 AU (20 and 280 mas). With only a single PFN baseline length, the exact location of such ring-shaped emission cannot be better constrained. In Section 4, we will compare to observations with previous $H$-band IOTA measurements of AB Aur with 10–40 m baselines, to try to further constrain the location of this hypothetical ring.

### 3.2. Azimuthal Signature

Several scenarios could explain the small ±0.2% anisotropic component (Figure 3).

3. An elliptical ring, or a circular ring, seen at a non-zero inclination angle (Figure 4(3)). In principle, it is possible to produce a small azimuthal null signature similar to that of Figure 3 with an elliptical ring. The best-fitting ($\chi^2$ of 2.2, Table 2) solution corresponds to an elliptical ring with about 6.5% of the total $K_s$-band flux with a semi-minor axis of 120 mas (17.3 AU), and semi-major axis of 200 mas (28.8 AU) at P.A. = 130° (Figure 6, blue dashed-triple-dotted curve). However, if originating from a circular ring seen at an angle, such a highly eccentric ellipse ($e = 0.73$) would correspond to an inclination of 53°, considerably larger than generally assumed for AB Aur (Eisner et al. 2003; Fukagawa et al. 2004). Moreover, such a large deviation from circularity seems inconsistent with the more generally circular structures seen at larger radii.

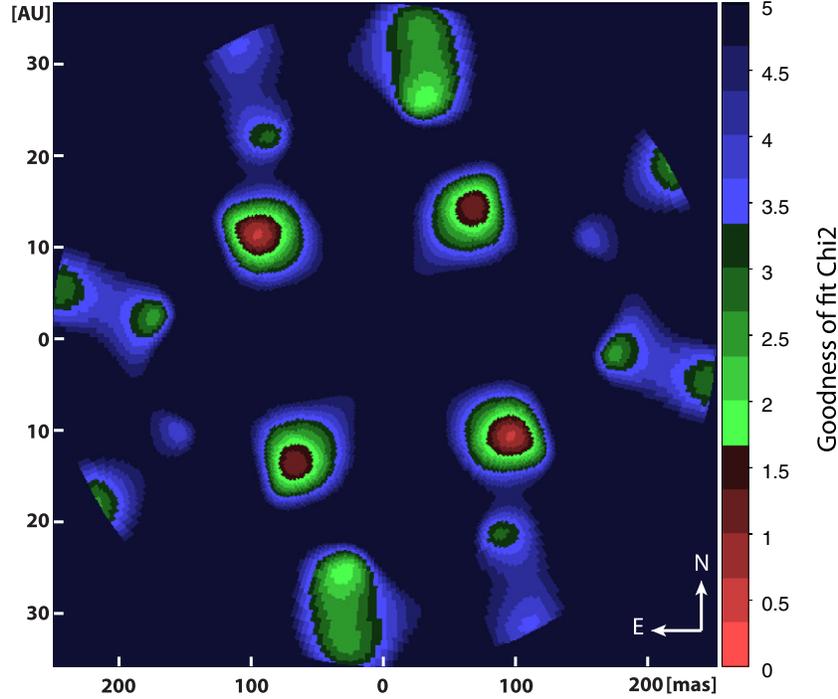

**Figure 7.** $\chi^2$ map for scenario #5 with an off-axis point source together with an azimuthally symmetric source responsible for the 1.35% pedestal null. The map shows the most probable location for a companion that could be responsible for the observed ±0.2% azimuthal modulation of Figure 3. Plotted field of view: 500 mas × 500 mas.

4. A spiral-like structure, possibly combined with isotropic emission (halo or ring). A simple model of thin spiral arms (Figure 4(4)) alone fails to reproduce the observed nulls of Figure 3, although it can reproduce the observed two oscillations per 180° rotation if the arms originate from a separation of about 100 mas (14.4 AU). However, one of the modulation peaks is always half the amplitude of the other, and in this case the amplitude of the predicted modulation is too large (±0.5% at least for a mean null of 1.5%, corresponding to 3.7% of the total $K_s$-band flux) for the spiral structure to be solely responsible for the measured PFN null signature. Nevertheless, combining the spiral structure with an extended halo is a potentially valid scenario to reproduce the measured astrophysical nulls: for example, a halo carrying about 4% of the total $K_s$-band flux combined with some elliptical structure accounting for about 1.25% of the flux, with a $\chi^2$ of about 2.7 (Table 2). However, even in this case, one of the modulation peaks is systematically half the amplitude of the other (Figure 6, yellow vertically elongated dashed curve).

5. An off-axis point source in addition to azimuthally isotropic emission (halo or ring), the latter being needed to explain the pedestal offset (Figure 4(5)). In order to explore this hypothesis, a 1.35% "pedestal null" (smallest measured null) is subtracted from the measured nulls of Figure 3. This would correspond to the signature of a slightly weaker azimuthally symmetric source (5.3% of the total $K_s$-band flux for a uniform halo), in order to leave room for the putative companion. One can then compute a $\chi^2$ map (Figure 7) measuring the probability that the remaining observed azimuthal null modulation is due to a point source at different locations around AB Aur, inside the PFN's FOV. The best fit is obtained for a companion at a separation of 122 ± 40 mas (17.6 ± 5.8 AU) and P.A. = 51° ± 13° *mod* 180°. As Figure 7 shows, solutions at the same separation but with P.A.s off by ±90° are also possible, but less likely. If the companion were located at the correlation peak of Figure 7, its flux would be 6.2 ± 0.25 × $10^{-3}$ of the unresolved central emission. This rather large flux could be indicative of a stellar mass companion or an accreting planet in formation (see scenario #6 below). Such a model actually provides the best possible fit to the observed on-sky nulls with the PFN ($\chi^2 = 0.6$ on Figure 6, red plain curve, as listed in Table 2).

6. A spatially extended off-axis source (a "blob"), such as a localized density enhancement, perhaps corresponding to, e.g., an accreting planet in formation (Figure 4(6), left). Such an emission source alone could, in principle, reproduce both the observed mean null level and azimuthal oscillation frequency (two periods per 180°), but to generate such a modest ±0.2% azimuthal modulation on top of a ~1.5% mean null, one would need to invoke large blob diameters, about 120 mas (17.3 AU), with a radial location of the same order, making such a model converge to a strongly decentered halo, which is hard to match with a physical model unless it is seen as leading to some extended spiral geometry (see model #4). Obviously, combining a spatially extended blob with an isotropic source such as a halo or ring (Figures 4(6), center and right) could also match the observed nulls (Figure 3), but such a model is inherently indeterminate, and is better covered by the scenario #5 of a point source in addition to a halo/ring as presented above. Indeed, given the rather large derived flux ($6.2 \times 10^{-3}$) for the best-fitting point-source location in Figure 7, an accreting planet in formation, with a blob-like geometry, is still a possibility.

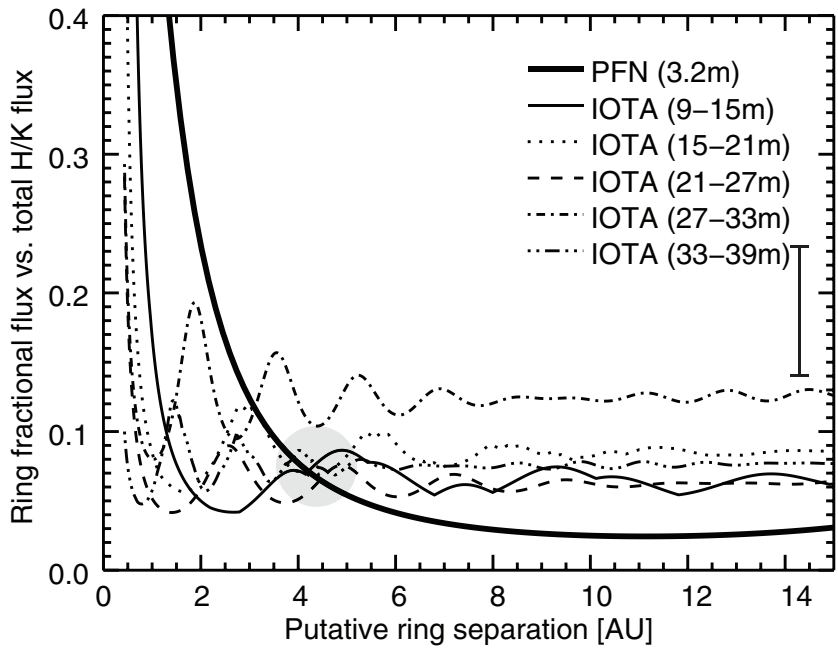

**Figure 8.** Derived constraints on the separation and flux of a hypothetical outer ring (model #2) vs. separation from *H*-band IOTA and $K_s$-band PFN data, solving Equation (1) for $F_{ring}(r)$ using previously reported models for the unresolved inner sub-AU components. All but one baseline groups intersect at a region located around 4.3 AU (30 mas), and for a ring contributing about 5%–10% of the total NIR flux (gray circle). The error bar on the right indicates the typical error on the IOTA-derived curves.

Overall, although the point-source hypothesis is one possible explanation for the observed azimuthal modulation, it is also possible that the origin of this slight anisotropic null signature is caused by some combination of an inclined structure (e.g., a ring) and disk asymmetry (e.g., spiral arms), perhaps similar to the ring-shaped structure at ∼40 AU (280 mas) recently imaged by polarimetry in *H* band (Hashimoto et al. 2011).

## 4. COMPARISON WITH PREVIOUS OBSERVATIONS

### 4.1. Compatibility of the PFN Outer Ring Model with Other Interferometric Observations

Comparing the PFN AB Aur data with previous longer baseline interferometric observations can further constrain the source of emission, in particular the components claimed to be partially resolved with the shortest baselines available at the time (∼9–10 m; Millan-Gabet et al. 2006b). As such, we examined 2006 AB Aur data obtained at IOTA (Traub et al. 2003) using the IONIC-3 *H*-band beam-combiner (Berger et al. 2003), covering a 9–40 m baseline range. Adopting the latest *H*- and *K*-band models for the inner sub-AU emission sources (primary, hot gas, and dusty ring accounting for 92%–95% of the total NIR excess; Millan-Gabet et al. 2006b; Tannirkulam et al. 2008), we tested the compatibility of the measured IOTA visibilities with a ring-shaped emission source (model #2) coming from outside 3 AU (21 mas), as suggested by the PFN measurements (Figure 5, solid line). Given that the IOTA data are in *H* band while our PFN measurements are in $K_s$ band, we only aim at rough compatibility estimates. Practically, the outer component suggested by IOTA and CHARA long baseline interferometric measurements is the only contribution potentially resolved by the PFN. The total flux of the sub-AU model—including the star itself—represents the PFN unresolved flux, and is the relevant normalizing factor for all measured PFN nulls. For an isotropic ring, we are in presence of a three-component system determining the measured interferometric fringe visibility $V_{obs}$ through the Van Cittert-Zernicke relation as follows:

$$V_{obs}(B) = \frac{F^* + [F_m(r) + F_{ring}(r)]V(B,r)}{F_{tot}}, \quad (1)$$

with $F_{tot} = F^* + F_m + F_{ring} = 1$ in this case (no incoherent emission, the PFN FOV being fully coherent), $F^*$ the unresolved star flux ratio over the total $K_s$-band emission (fraction of total light), $F_m(r)$ the model inner sub-AU component flux ratio as a function of the distance to the star $r$, $F_{ring}(r)$ the flux of a putative thin ring with a unknown location and flux ratio, and $B$ the baseline length. All components but the unresolved star are modulated by the known radially dependent fringe visibility function $V(B, r)$, taking finite coherent FOV effects into account.

By inserting a sub-AU scale model $F_m(r)$ in Equation (1), such as the ones derived by Millan-Gabet et al. (2006b) and Tannirkulam et al. (2008), as well as the measured visibilities $V_{obs}$ from the IOTA data set, we can solve Equation (1) for the flux of the model #2 "missing ring" $F_{ring}$ in function of stellocentric separation $r$, and this for each baseline $B$ probed.

Figure 8 shows for different baselines (IOTA and PFN), the relative flux $F_{ring}$ of a thin ring located at various distances from the central star. It is computed from Equation (1) using all the IOTA measured visibilities and their error bars. The retained *H*-band model for the inner sub-AU component percentage of the total flux is the following: a star flux ratio of $F^* = 30\%$ and a hot Gaussian dust ring model with total flux $F_m = 62\%$ (ring radius = 1.75 mas, ring width/diameter = 0.5). This corresponds to the previous *H*-band model from Millan-Gabet et al. (2006b), revised with the new ring separation estimate slightly further out at 0.25 AU (1.75 mas) from Tannirkulam et al. (2008). Given that more than 80 baselines with various azimuthal projections on-sky were recorded in the 2006 IOTA data set, but with no clear azimuthal signatures and relatively large error bars (Millan-Gabet et al. 2006b), the $F_{\rm ring}$ values plotted in Figure 8 are binned over five baseline intervals covering the full 9–39 m range for increased clarity purpose. Also plotted in Figure 8 is the ring flux curve derived from the PFN data (same as Figure 5).

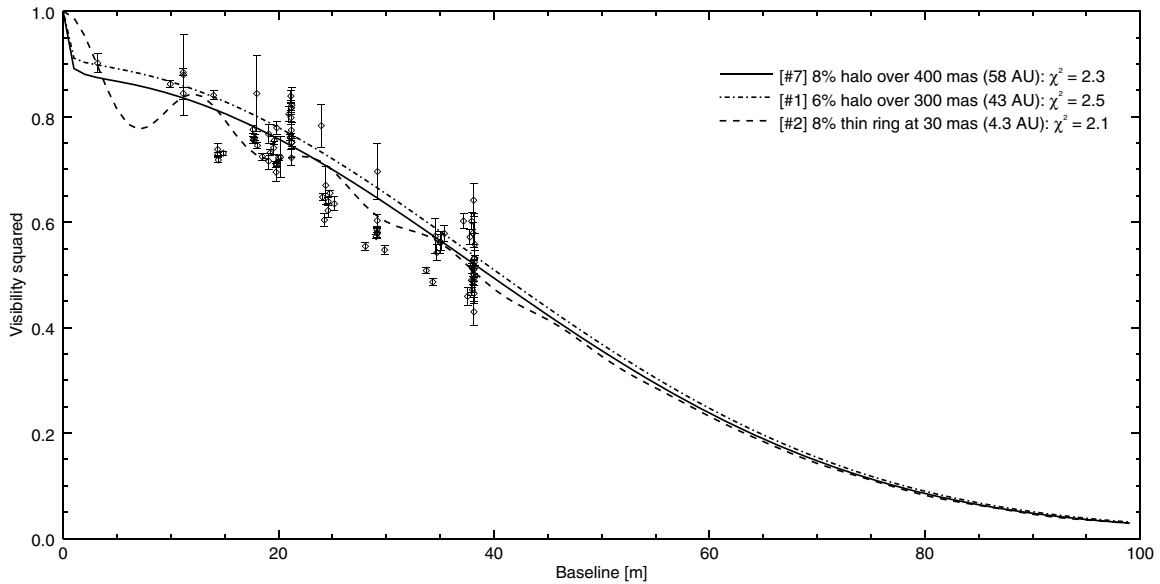

**Figure 9.** 2006 IOTA *H*-band and 2012 PFN $K_s$-band measured squared visibilities squared plotted with best-fit azimuthally symmetric models. When computing the various models' $\chi^2$ values, measurements obtained at different azimuths, but for very similar baseline lengths (within 1 m), are binned together, in order to guarantee that no baseline length is strongly favored.

In spite of relatively large error bars coming from the 2006 IOTA visibility measurements and the difference in wavelength, the plot shows an intersection region between the PFN curve and all but one IOTA curve (gray circle in Figure 8). This common solution corresponds to a ring carrying 5%–10% of the total NIR flux and located at about 30 mas (4.3 AU). Interestingly, Leinert et al. (2001) quoted the same separation of 30 mas as the characteristic NIR size for AB Aur, using speckle interferometry at *K* band to probe spatial scales beyond 200 mas (29 AU). Secondly, Millan-Gabet et al. (2006b) also suggested the presence of a source at a separation of 30 mas, contributing about 8% of the total *H*-band flux. However, this source was suggested to be a 1.7 AU wide (12 mas) "blob" (potentially an accreting planet in formation) and was not as spatially extended as the ring model considered here.

In any case, 30 mas (4.3 AU) seems to be a characteristic size scale for spatially extended models. Specifically, we will verify the compatibility of model #2 with a ring at 30 mas (4.3 AU) with the measured visibilities at IOTA by adding a new visibility data point corresponding to the PFN's 3.2 m baseline (Figure 9) to the IOTA visibility data. The PFN visibility value is computed from the mean measured astrophysical null $N_{\rm as}$ (Figure 3) as $V_{\rm obs,\,PFN} \sim (1 - N_{\rm as})/(1 + N_{\rm as})$ and adjusted for the IOTA transmission inside the same FOV. The model #2 with an 8% additional ring of dust at 30 mas (dashed curve in Figure 9) is combined with previously determined components inside of 2 mas (central star, gas, and inner ring; Millan-Gabet et al. 2006b; Tannirkulam et al. 2008). Also plotted in Figure 9 are two variants of the extended halo model:

*Model #7.* An 8% halo filling the entire IOTA 400 mas (58 AU) FOV as in Millan-Gabet et al. (2006b) and Monnier et al. (2006; solid curve).

*Model #1.* A 6% halo filling the entire PFN 300 mas (43 AU) FOV (dashed-point curve).

As can be in seen in Figure 9, by introducing oscillations in the visibility curve, the thin outer ring #2 model can better reproduce some features of the 2006 IOTA visibility measurements, and provides a marginally better fit ($\chi^2 = 2.1$) than the #7 IOTA ($\chi^2 = 2.3$) and #1 PFN ($\chi^2 = 2.5$) extended halo models. The presence of an outer emission ring at $\sim$30 mas (4.3 AU) and carrying $\sim$8% of the central flux, which provides a good fit to the PFN data, is clearly also compatible with all interferometric results obtained for AB Aur.

### 4.2. Compatibility of Previously Suggested Models with the PFN Results

Conversely, models previously suggested by other authors can be tested for compatibility with the PFN observations, taking into account the expected large orbital motion between observational epochs at these probed scales. We studied the compatibility of four other models, three attempting to explain the "2%–8% large-scale NIR excess," as proposed previously to explain the visibility deficit observed at relatively short baselines ($<$20 m). The fourth one is a hybrid adaptation from the mid-IR model of Monnier et al. (2009). They consist of the following.

6. An 8% 12 mas (1.7 AU) wide Gaussian "blob" at 29 mas (4.2 AU), P.A. $\sim 12°$ (Millan-Gabet et al. 2006b). We forward-propagated this body along a circular orbit from its 2003 IOTA epoch to the 2012 PFN observations, yielding an October 2012 P.A. of $\sim$205°.
7. An 8% so-called uniform halo model extending up to 400 mas (57.6 AU, the incoherent FOV for IOTA; Monnier et al. 2006; Tannirkulam et al. 2008).
8. A hypothetical 2% close-in point source at 9 mas (1.3 AU), P.A. $\sim 22°$ (Millan-Gabet et al. 2006b).
9. A halo, with some slight extension along P.A. $\sim 27°$, similarly to the reported geometry at 10 $\mu$m with Keck segment tiling observations (Monnier et al. 2009). The large ($\sim$89% light fraction) reported mid-IR characteristic Gaussian emission at 7 AU (35 mas) FWHM scales is entirely incompatible with the detected mean astrophysical null of just 1.52%, as it would have resulted in a null leakage of about 25%. However, we scaled down in brightness the Gaussian spatial model accordingly, and converted the reported slight "extension" at these scales by forward-propagation from its 2004 August Keck epoch to our 2012 October PFN epoch.

The four corresponding brightness distributions (#6 through #9, see Table 2) were multiplied by the PFN transmission profile of Figure 1. The expected azimuthal null responses are plotted in Figure 6, together with previous models derived from the PFN observations.

From Figure 6 and the large corresponding $\chi^2$ values (Table 2), it appears that models #8 ($\chi^2 = 395$) and #6 ($\chi^2 = 471$) are not compatible with our PFN observations, at least not as the sole relevant (>1%) source of the NIR excess detected past 1 AU (7 mas). Even combining the reported #8 point-source model with any spatially extended symmetric distribution would not match the observed nulls in terms of azimuthal signature, which is indicative of a larger separation from the primary if this companion exists, such as model #5. It is, however, worth noting that even if its radial separation and putative flux does not match the measured "double-peaked" PFN response on-sky due to it being too close, the Gaussian "blob" #6 model is just within the error bars of the P.A. ∼ 51 mod 180° for the 0.6% putative point-source #5 model embedded in an extended structure (Figure 7), once forward-propagated to a 2012 epoch (new P.A. of 205° E of N).

The scaled-down mid-IR #9 model does not match the PFN measurements either in Figure 6 ($\chi^2 = 9.4$), but such a discrepancy is not inherently surprising given that the $K_s$-band PFN observation looks at much hotter dust than the cold contribution at mid-IR wavelengths.

Finally, an extended halo model accounting for 8% of the $K$-band flux (model #7, long dashed curve in Figure 6) and homogenously located inside IOTA FOV (400 mas inward) is not large enough to explain the detected excess ($\chi^2 = 60$, Table 2). A similarly extended lower halo flux at $K$ band compared to $H$ band, e.g., 5% as in Tannirkulam et al. (2008), would match the PFN observations even less.

Overall, there is therefore a reasonable rationale to assume that most—if not all—of any NIR halo or "missing" flux reported by long baseline interferometry falls inside the 4–40 AU PFN FOV, and the three main reasons are the following: most NIR long baseline interferometry observations agree that about 95% of the $H$-/$K$-band flux is located inside 1 AU (7 mas), leaving ∼5% of incoherent flux located somewhere farther out (Millan-Gabet et al. 2006b; Monnier et al. 2006; Tannirkulam et al. 2008), $K$-band speckle interferometry observations by Leinert et al. (2001) largely exclude any significant emission on scales beyond 29 AU (200 mas), and the ∼1.5% null leakage detected at all azimuths by the PFN already corresponds to a spatially extended emission of about 5%–10% of the $K_s$-band flux inside the PFN's 40 AU (275 mas) FOV, for both a halo or best-fit ring model, so that no additional outside flux is required to reproduce the data.

## 5. CONCLUSIONS AND PERSPECTIVES

The NIR PFN measurements uniquely explore intermediate scales corresponding to the outer solar system planetary region (5–40 AU) around the young pre-main-sequence star AB Aur, i.e., those between earlier NIR long baseline interferometry sub-AU observations, and >40 AU direct imaging observations. Within this range of physical distances, we detected an azimuthally extended source contributing ∼5%–10% of the total $K_s$-band (∼2.2 $\mu$m) flux of AB Aur. Combining our PFN results with previous IOTA observations at $H$ band, our best-fit model is an ∼8% ring of dust located at 4 AU (30 mas) from the star. A ring enhancement might presumably also imply the presence of an emission deficit in the intermediate region between ∼0.5 and 4 AU. Such "gapped-disk" geometries have indeed been recently suggested as a relevant model for "flaring" Herbig Ae/Be stars, based on mid-IR spectral energy distribution (SED) data (Maaskant et al. 2013) and have been seen in great details by ALMA around HL Tau. However, a reasonably good fit can also be obtained assuming a ∼6% uniform "halo" extending up to 43 AU (300 mas).

Furthermore, we also measured a ±0.2% azimuthal variation of the null, which could reflect some skewness or anisotropy in the extended source of emission resolved by the PFN, such as, e.g., spiral arm(s). Alternatively, this small azimuthal signature could also trace the presence of an embedded point-source companion, with a fractional flux of $6 \pm 0.25 \times 10^{-3}$ and located at $17.6 \pm 5.8$ AU and P.A. = $52° \pm 13°$ mod $180°$. It could also be caused by a 6.5% elliptical ring with axis ranging from 17 to 29 AU (equivalent to a warped 29 AU radius circular ring seen at a 53° inclination). On the other hand, several previously hypothesized models seem incompatible. Clearly, given the rather significant NIR emission detected inside the PFN 5–40 AU (35–280 mas) FOV, there is a significant rationale for concluding that the detected contribution should correspond to most of the "large-scale halo" flux previously reported for AB Aur with NIR long baseline interferometry (Monnier et al. 2009; Tannirkulam et al. 2008), leaving little significant emission to look for at larger separations. This highlights a promising survey capability of the PFN, enabling one to put new strong constrains on the location of the "large-scale halos" found around YSOs, notably around FU-Orionis objects (Leinert et al. 2001; Millan-Gabet et al. 2006a). The origin of such a spatially extended warm emission at this range of separations (past 4 AU) is inherently intriguing, as it is unlikely to be thermal emission. Scattering seems to be a more plausible explanation, though one would need to find a way to illuminate the dust in a nearly isotropic way despite the shadow cast by the inner hot dust rim. In this context, we note that the disk-wind class of models (e.g., Bans & Königl et al. 2012) could provide a possible pathway for this, by enabling the dust to be lifted up above the main disk plane past a few multiples of the sublimation radius, hence potentially ending up being illuminated by the central stellar component. Here again, the PFN can probably provide clues on the validity of these models and refine their geometric parameters constraints.

Follow-up observations of AB Aur with different PFN baseline lengths (e.g., 2.6–3.8 m) and at different epochs should help to discriminate between these models. Typically, it will be very valuable to monitor the evolution of the faint ±0.2% azimuthal null modulation, in order to either discard the point-source hypothesis or start to put together orbital parameters. More generally, future high angular resolution direct coronagraphic imaging observations should begin to provide better overlap with the smaller PFN FOV in the 10–40 AU range, allowing better definition of source structures, discriminating between dust structures and companions. Finally, the implementation of a rotating baseline nuller on a larger telescope will be able to push into nearly AU scales even at the 140 pc distance of AB Aur.

This work was performed at the Jet Propulsion Laboratory, California Institute of Technology, under contract with NASA. The data presented are based on observations obtained at the Hale Telescope, Palomar Observatory, as part of a continuing collaboration between Caltech, NASA/JPL, and Cornell University. We particularly thank the staff of the Palomar Observatory for their assistance in mounting the PFN and in conducting the observations at the Hale telescope. J.K. is supported by a Swiss National Science Foundation Advanced Postdoc Mobility fellowship (PA00 P2 136416). We also thank Rens Waters for the helpful discussions.